\documentstyle[ltwol]{article}

\input{psfig}

\def\ijmp#1#2#3{{\it Int.\ J.\ Mod.\ Phys.} A {\bf #1}, #3 (19#2)}
\def\newprd#1#2#3{{\it Phys.\ Rev.} D {\bf #1}: #3 (19#2)}
\def\newprdtwo#1#2#3{{\it Phys.\ Rev.} D {\bf #1}: #3 (20#2)}
\def\plb#1#2#3{{\it Phys.\ Lett.} B {\bf #1}, #3 (19#2)}
\def\prd#1#2#3{{\it Phys.\ Rev.} D {\bf #1}, #3 (19#2)}
\def\prl#1#2#3{{\it Phys.\ Rev.\ Lett.} {\bf #1}, #3 (19#2)}
\def\zpc#1#2#3{{\it Zeit.\ Phys.} C {\bf #1}, #3 (19#2)}
\def\npb#1#2#3{{\it Nucl.\ Phys.} B {\bf #1}, #3 (19#2)}

\def\beq{\begin{equation}}
\def\eeq{\end{equation}}
\def\bea{\begin{eqnarray}}
\def\eea{\end{eqnarray}}
\def\nn{\nonumber}
\def\sss{\scriptscriptstyle}

\def\bd{B_d^0}
\def\bdbar{{\overline{B_d^0}}}
\def\bs{B_s^0}
\def\bsbar{{\overline{B_s^0}}}
\def\barp{{\raise.35ex\hbox
{${\sss (}$}}---{\raise.35ex\hbox{${\sss )}$}}}
\def\bdbarp{\hbox{$B_d$\kern-1.4em\raise1.4ex\hbox{\barp}}}
\def\bsbarp{\hbox{$B_s$\kern-1.4em\raise1.4ex\hbox{\barp}}}
\def\ks{K_{\sss S}}
\def\kbar{{\overline{K^0}}}
\def\roughly#1{\mathrel{\raise.3ex\hbox
{$#1$\kern-.75em\lower1ex\hbox{$\sim$}}}}

\def\Abar{\bar A}

\def\tnp{\theta_{\sss NP}}

\def\adirpm{{a_{\sss dir}^{+-}}}
\def\adir00{{a_{\sss dir}^{00}}}

\def\alphaeff{{\alpha_{\sss eff}}}

\def\alphaeffpm{{\alpha_{\sss eff}^{+-}}}

\def\B00{B^{00}}
\def\Bp0{B^{+0}}

\newread\epsffilein 
\newif\ifepsffileok 
\newif\ifepsfbbfound 
\newif\ifepsfverbose 
\newdimen\epsfxsize 
\newdimen\epsfysize 
\newdimen\epsftsize 
\newdimen\epsfrsize 
\newdimen\epsftmp 
\newdimen\pspoints 
\def\epsfbox#1{\global\def\epsfllx{72}\global\def\epsflly{72}%
 \global\def\epsfurx{540}\global\def\epsfury{720}%
 \def\lbracket{[}\def\testit{#1}\ifx\testit\lbracket
 \let\next=\epsfgetlitbb\else\let\next=\epsfnormal\fi\next{#1}}%
\def\epsfgetlitbb#1#2 #3 #4 #5]#6{\epsfgrab #2 #3 #4 #5 .\\%
 \epsfsetgraph{#6}}%
\def\epsfnormal#1{\epsfgetbb{#1}\epsfsetgraph{#1}}%
\def\epsfgetbb#1{%
\openin\epsffilein=#1
\ifeof\epsffilein\errmessage{I couldn't open #1, will ignore it}\else
 {\epsffileoktrue \chardef\other=12
 \def\do##1{\catcode`##1=\other}\dospecials \catcode`\ =10
 \loop
 \read\epsffilein to \epsffileline
 \ifeof\epsffilein\epsffileokfalse\else
 \expandafter\epsfaux\epsffileline:. \\%
 \fi
 \ifepsffileok\repeat
 \ifepsfbbfound\else
 \ifepsfverbose\message{No bounding box comment in #1; using defaults}\fi\fi
 }\closein\epsffilein\fi}%
\def\epsfclipstring{}
\def\epsfsetgraph#1{%
 \epsfrsize=\epsfury\pspoints
 \advance\epsfrsize by-\epsflly\pspoints
 \epsftsize=\epsfurx\pspoints
 \advance\epsftsize by-\epsfllx\pspoints
 \epsfxsize\epsfsize\epsftsize\epsfrsize
 \ifnum\epsfxsize=0 \ifnum\epsfysize=0
 \epsfxsize=\epsftsize \epsfysize=\epsfrsize
 \epsfrsize=0pt
 \else\epsftmp=\epsftsize \divide\epsftmp\epsfrsize
 \epsfxsize=\epsfysize \multiply\epsfxsize\epsftmp
 \multiply\epsftmp\epsfrsize \advance\epsftsize-\epsftmp
 \epsftmp=\epsfysize
 \loop \advance\epsftsize\epsftsize \divide\epsftmp 2
 \ifnum\epsftmp>0
 \ifnum\epsftsize<\epsfrsize\else
 \advance\epsftsize-\epsfrsize \advance\epsfxsize\epsftmp \fi
 \repeat
 \epsfrsize=0pt
 \fi
 \else \ifnum\epsfysize=0
 \epsftmp=\epsfrsize \divide\epsftmp\epsftsize
 \epsfysize=\epsfxsize \multiply\epsfysize\epsftmp
 \multiply\epsftmp\epsftsize \advance\epsfrsize-\epsftmp
 \epsftmp=\epsfxsize
 \loop \advance\epsfrsize\epsfrsize \divide\epsftmp 2
 \ifnum\epsftmp>0
 \ifnum\epsfrsize<\epsftsize\else
 \advance\epsfrsize-\epsftsize \advance\epsfysize\epsftmp \fi
 \repeat
 \epsfrsize=0pt
 \else
 \epsfrsize=\epsfysize
 \fi
 \fi
 \ifepsfverbose\message{#1: width=\the\epsfxsize, height=\the\epsfysize}\fi
 \epsftmp=10\epsfxsize \divide\epsftmp\pspoints
 \vbox to\epsfysize{\vfil\hbox to\epsfxsize{%
 \ifnum\epsfrsize=0\relax
 \includegraphics{#1}%
 \else
 \epsfrsize=10\epsfysize \divide\epsfrsize\pspoints
 \includegraphics{#1}%
 \fi
 \hfil}}%
\global\epsfxsize=0pt\global\epsfysize=0pt}%
 {\catcode`\%=12 \global\let\epsfpercent=
\long\def\epsfaux#1#2:#3\\{\ifx#1\epsfpercent
 \def\testit{#2}\ifx\testit\epsfbblit
 \epsfgrab #3 . . . \\%
 \epsffileokfalse
 \global\epsfbbfoundtrue
 \fi\else\ifx#1\par\else\epsffileokfalse\fi\fi}%
\def\epsfempty{}%
\def\epsfgrab #1 #2 #3 #4 #5\\{%
\global\def\epsfllx{#1}\ifx\epsfllx\epsfempty
 \epsfgrab #2 #3 #4 #5 .\\\else
 \global\def\epsflly{#2}%
 \global\def\epsfurx{#3}\global\def\epsfury{#4}\fi}%
\def\epsfsize#1#2{\epsfxsize}
\begin{document}

\twocolumn[

\begin{flushright}
UdeM-GPP-TH-01-82 \\
\end{flushright}

\vskip0.3truecm

\centerline{\bf Searching for New Physics via CP Violation in
  $B\to\pi\pi$}

\vskip0.5truecm

\centerline{David London}

\vskip0.2truecm

\centerline{\it Laboratoire Ren\'e J.-A. L\'evesque, Universit\'e de
  Montr\'eal,}
\centerline{\it C.P. 6128, succ. centre-ville, Montr\'eal, QC, Canada
  H3C 3J7}
\centerline{\it email: london@lps.umontreal.ca}

\vskip0.3truecm

\centerline{Nita Sinha and Rahul Sinha}

\vskip0.2truecm

\centerline{\it Institute of Mathematical Sciences, Taramani, Chennai 600113,
  India}
\centerline{\it email: nita@imsc.ernet.in , sinha@imsc.ernet.in}


\vskip0.5truecm

\abstracts{ We show how $B\to\pi\pi$ decays can
  be used to search for new physics in the $b\to d$ flavour-changing
  neutral current. One needs one piece of theoretical input, which we
  take to be a prediction for $P/T$, the ratio of the penguin and tree
  amplitudes in $\bd\to\pi^+\pi^-$. If present, new physics can be
  detected over most of the parameter space. If $\alpha$ ($\phi_2$)
  can be obtained independently, measurements of $B^+ \to \pi^+ \pi^0$
  and $\bd/\bdbar \to \pi^0\pi^0$ are not even needed. }]

$B$-factory measurements have provided us with the first hints for CP
violation in the $B$ system. We all expect that, with time, it will be
possible to measure CP violation in a wide variety of $B$ decays. And
when the dust settles, we all hope that these measurements will reveal
the presence of physics beyond the standard model (SM).

New physics can alter CP-violating rate asymmetries in $B$ decays
principally by affecting those amplitudes which in the SM are at the
one-loop level.~\cite{NPBmixing} These new effects can be separated
into two classes. New physics can affect the $b\to d$ flavour-changing
neutral current (FCNC), which includes $\bd$-$\bdbar$ mixing and $b\to
d$ penguin amplitudes. It can also affect the $b\to s$ FCNC
($\bs$-$\bsbar$ mixing, $b\to s$ penguin amplitudes).

There are several clean, direct tests for new physics in the $b\to s$
FCNC. For example, the decays $B^\pm \to D K^\pm$~\cite{BDK} and
$\bs(t) \to D_s^\pm K^\mp$~\cite{BsDsK} both probe the CP phase
$\gamma$ ($\phi_3$) in the SM.  A discrepancy between these two CP
asymmetries will point clearly to the presence of new physics in
$\bs$-$\bsbar$ mixing. Similarly, if it is found that the CP
asymmetries in $\bd(t) \to \psi\ks$ and $\bd(t) \to \phi\ks$ are
unequal (they both measure $\beta$ ($\phi_1$) in the SM), this will
indicate new physics in the $b\to s$ penguin amplitude.~\cite{LonSoni}
Finally, the decay $\bs(t) \to \psi\phi$ is expected to have a tiny CP
asymmetry in the SM. If this turns out not to be the case, we will
know that there is new physics in $\bs$-$\bsbar$ mixing.

This then begs the question: are there clean, direct tests for new
physics in the $b\to d$ FCNC? At first glance, the answer appears to
be `yes'. Assuming that the $b\to d$ penguin amplitude is dominated by
the exchange of an internal $t$ quark, its weak phase is $-\beta$ in
the Wolfenstein parametrization.~\cite{Wolfenstein} One then predicts
that the CP asymmetry in the decay $\bd(t) \to K^0\kbar$ vanishes,
while $\bs(t)\to\phi\ks$ measures $\sin 2\beta$.~\cite{LonPeccei} Any
deviation from these predictions would indicate new physics in the
$b\to d$ FCNC.

Unfortunately, the assumption of $t$-quark dominance of the $b\to d$
penguin amplitude is incorrect. The $u$- and $c$-quark contributions
can be quite substantial, as large as 20--50\% of the $t$-quark
contribution.~\cite{ucquark} Thus, the weak phase of the $b\to d$
penguin is {\it not} $-\beta$, and the clean predictions for the
asymmetries in $\bd(t) \to K^0\kbar$ and $\bs(t)\to\phi\ks$ are
spoiled.

But this then raises a second question: can one isolate the $t$-quark
contribution to the $b\to d$ penguin, and measure its weak phase? If
so, then the comparison of this weak phase with that measured in
$\bd(t) \to\psi\ks$ could reveal the presence of new physics.
Unfortunately, as shown in Ref.~8, the answer to this question is
`no'. It is not possible to isolate any single contribution to the
$b\to d$ penguin. Thus, it is impossible to cleanly test for new
physics in the $b\to d$ FCNC.

However, one {\it can} test for new physics if we make a single
assumption about the theoretical parameters describing the
decay.~\cite{CKMambiguity} In this talk we describe how this can be
applied to $B\to \pi\pi$, which, as is well known, suffers from
penguin ``pollution.'' As we will see, the measurements of $B\to
\pi\pi$, combined with a theoretical prediction for $P/T$, the ratio
of penguin and tree amplitudes in $\bd\to\pi^+\pi^-$, allow one to
probe new physics in the $b\to d$ penguin amplitude.~\cite{Bpipi}

We begin with a brief review of $B\to \pi\pi$ decays. Recall that, in
the Wolfenstein parametrization, the weak phase of $\bd$-$\bdbar$
mixing is $-2\beta$. It is convenient to remove this phase by
redefining the $B$ decay amplitudes:
\beq 
A^f \equiv e^{i \beta} Amp(\bd \to f) ~~,~~~~ \Abar^f \equiv e^{-i\beta}
Amp({\bar \bd} \to f) ~.
\eeq
With this convention, the time-dependent decay rate for $\bd(t) \to
\pi^+ \pi^-$ takes the form
%
\bea
\Gamma(\bd(t) \to \pi^+\pi^-) & = & e^{-\Gamma t} \left[ {|A^{+-}|^2 +
    |\Abar^{+-}|^2 \over 2} \right. \nn\\
& & + {|A^{+-}|^2 - |\Abar^{+-}|^2 \over 2}
  \cos (\Delta M t) \nn\\
& & \left. - {\rm Im} \left({A^{+-}}^* \Abar^{+-}\right) \sin
  (\Delta M t) \right] .
\label{timedep}
\eea
Thus, the measurement of the time-dependent decay rate $\bd(t) \to
\pi^+ \pi^-$ allows the extraction of the following three quantities:
\bea 
B^{+-} &\equiv& {1\over 2} \left( |A^{+-}|^2+|{\bar A^{+-}}|^2 \right) ~,~\nn\\
\adirpm &\equiv& {{|A^{+-}|^2-|{\bar A^{+-}}|^2} \over
{|A^{+-}|^2+|{\bar A^{+-}}|^2} }~,~\nn\\
2\alphaeffpm &\equiv&
Arg\left({A^{+-}}^* \Abar^{+-}\right) ~.
\eea
In what follows, the direct CP asymmetry $\adirpm$ and the indirect CP
asymmetry $2\alphaeffpm $ will be the key observables. Similarly, one
can also obtain $B^{00}$ and $\adir00$ through measurements of
$\bd/\bdbar \to \pi^0\pi^0$, and $B^{+0}$ from $B^+ \to \pi^+ \pi^0$.

In addition to a tree amplitude, the decay $\bd\to\pi^+\pi^-$ also
receives a contribution from a $b\to d$ penguin amplitude. Eliminating
the $c$-quark piece of the penguin amplitude ($V_{cb}^* V_{cd}$) using
the unitarity of the CKM matrix, in the SM one can write
\beq
A^{+-} = T e^{i\delta} e^{-i\alpha} + P e^{i \delta_{\sss P}} ~.
\eeq
Here $\alpha$ is one of the three angles of the unitarity triangle,
$\delta$ and $\delta_{\sss P}$ are strong phases, and $P$ and $T$ are
defined to be real, positive quantities. Note that the $T$ amplitude
is not pure tree: it includes contributions from the $u$- and
$c$-quark pieces of the $b\to d$ penguin. Similarly, the $P$ amplitude
is mostly the $t$-quark penguin contribution, but includes a $c$-quark
penguin piece. Note also that there is no weak phase multiplying the
$P$ amplitude. This is because, in the SM, the weak phase of
$\bd$-$\bdbar$ mixing cancels that of the $t$-quark piece of the $b\to
d$ penguin amplitude.

Now, it is well known that the isospin analysis of $B\to\pi\pi$ decays
enables one to remove the penguin pollution and extract the CP phase
$\alpha$.~\cite{isospin} However, it is also true, though not as well
known, that this same analysis allows one to obtain all of the
theoretical parameters $T$, $P$, etc.~\cite{Charles} In particular,
\beq
r^2 \equiv \frac {P^2}{T^2} = {1-\sqrt{1 - (\adirpm)^2} \cos(2 \alpha
  - 2 \alphaeff) \over 1-\sqrt{1 - (\adirpm)^2} \cos(2 \alphaeff)} ~.
\eeq 
If there is new physics in the $b\to d$ penguin amplitude, the
$\bd\to\pi^+\pi^-$ amplitude will be modified:
\beq
A^{+-} = T e^{i\delta} e^{-i\alpha} + P e^{i \delta_{\sss P}} e^{-i\tnp} ~,
\eeq
where $\tnp$ represents the mismatch, due to the presence of new
physics, between the weak phase of $\bd$-$\bdbar$ mixing and that of
the $t$-quark piece of the $b\to d$ penguin amplitude. The expression
for $r^2$ now reads
\beq
r^2 \equiv \frac {P^2}{T^2} = {1-\sqrt{1 - (\adirpm)^2} \cos(2 \alpha
  - 2 \alphaeff) \over 1-\sqrt{1 - (\adirpm)^2} \cos(2\tnp - 2 \alphaeff)} ~.
\label{r}
\eeq 
{}From this expression it is clear that, given measurements of
$\adirpm$, $2\alphaeff$ and $2\alpha$, along with a theoretical
prediction of $P/T$, one can extract $\tnp$. (Note that, in the above,
we have assumed that new physics affects only the weak phase of the
$P$ piece of $A^{+-}$. However, new physics could also affect the
magnitude of $P/T$. For our purposes, since we are looking for
$\tnp\ne 0$, this distinction is unimportant: if, in reality, $\tnp=0$
but new physics has affected the magnitudes of $P$ and $T$, this will
still show up as an effective nonzero $\tnp$.)

Of course, in practice, things are more complicated. First, theory
will not predict a specific value for $P/T$, but rather a range of
values. For example, Fleischer and Mannel give $0.07 \le r \le
0.23$,~\cite{FM} while Gronau estimates $r = 0.3 \pm
0.1$.~\cite{Gronau} In this study, we will take a very conservative
range:
\beq
0.05 \leq r \leq 0.5 ~.
\label{rbound}
\eeq

Second, the formula for $P/T$ depends on $2\alpha$. Where does one get
this CP phase? Obviously, if the isospin analysis can be performed,
one can obtain it in that way. However, it may be very difficult to
perform such an analysis, in which case we will need to get $2\alpha$
from outside of the $B\to\pi\pi$ system. One possibility is to measure
$2\alpha$ using the Dalitz-plot analysis of $B\to\rho\pi$
decays.~\cite{Dalitz} Another possibility, assuming that both $\beta$
and $\gamma$ have been measured, is to use the relation $\alpha = \pi
- \beta - \gamma$, which holds even in the presence of new
physics.~\cite{nirsilv} The bottom line is that there are a number of
ways of getting $2\alpha$, and ideally we will have information from
all of these sources.

The third complication is the fact that all measurements will be made
with some error, and these errors can mask the presence of a nonzero
$\tnp$. To approximate this effect, we take $2\alpha$ to lie in a
certain range, and we consider two such illustrative choices: 
\bea
& & (a) ~~ 120^\circ \le 2\alpha \le 135^\circ, \nn\\
& & (b) ~~ 165^\circ \le 2\alpha \le 180^\circ.
\label{alphachoices}
\eea
The procedure is now nominally as follows: given measurements of
$\adirpm$ and $2\alphaeff$, and assuming that $2\alpha$ lies in range
(a) or (b), we will see if $\tnp = 0$ gives $r$ in the allowed range
[Eq.~(\ref{rbound})].  If not, this indicates the presence of new
physics.

However, this does not take into account all information we have at
our disposal. By the time $\bd(t) \to \pi^+\pi^-$ is measured, we will
have some knowledge of the branching ratios for $B^+ \to \pi^+ \pi^0$
and $\bd/\bdbar \to \pi^0\pi^0$. (Indeed, even today we already have
upper limits on these quantities.~\cite{Gilman}) Thus, we can use
isospin to test for new physics. That is, given a range of values for
$B^{+0}$, $B^{00}$ and $\adir00$, it must be possible to reproduce the
measured value of $2\alpha$ using the isospin analysis. If not, this
indicates the presence of new physics.

To include this constraint, we consider five scenarios for the allowed
ranges of $B^{+0}/B^{+-}$, $B^{00}/B^{+-}$ and $\adir00$, shown in
Table 1. Note that scenario A assumes that we have no knowledge of
these quantities at all. Obviously this is totally unrealistic, since
we already have upper limits on $B^{+0}/B^{+-}$ and $B^{00}/B^{+-}$. A
more realistic case, which is roughly the situation as it exists
today, is given in scenario B. Here, $B^{+0}$ is approximately equal
to $B^{+-}$, while there is an upper limit on $B^{00}/B^{+-}$ and
nothing is known about $\adir00$. Finally, scenarios C, D and E are
other hypothetical ranges for $B^{+0}/B^{+-}$, $B^{00}/B^{+-}$ and
$\adir00$.

\begin{table}
\hfil
\vbox{\offinterlineskip
\halign{&\vrule#&
 \strut\quad#\hfil\quad\cr
\noalign{\hrule}
height2pt&\omit&&\omit&&\omit&&\omit&\cr
& \omit && $\adir00$ && $B^{00}/B^{+-}$ && $B^{+0}/B^{+-}$ & \cr
height2pt&\omit&&\omit&&\omit&&\omit&\cr
\noalign{\hrule}
height2pt&\omit&&\omit&&\omit&&\omit&\cr
& Case A && $-1$ -- 1 && any value && any value & \cr
& Case B && $-1$ -- 1 && 0 -- 0.1 && 0.8 -- 0.9 & \cr
& Case C && 0.5 -- 0.7 && 0.7 -- 0.8 &&  0 -- 0.5 & \cr
& Case D && 0.6 -- 1 && 0.2 -- 0.4 && 0.6 -- 0.7 & \cr
& Case E && 0.6 -- 1 && 0.2 -- 0.4 && 0.2 -- 0.3 & \cr
height2pt&\omit&&\omit&&\omit&&\omit&\cr
\noalign{\hrule}}}
\caption{The assumed ranges for $\adir00$, $B^{00}/B^{+-}$ and 
$B^{+0}/B^{+-}$ for five (hypothetical) sets of experimental 
measurements.}
\label{casetable}
\end{table}

In order to map out the region of parameter space where new physics
can be found, we use the following procedure. In a given scenario, we
generate values for $\adirpm$ and $2\alphaeff$ in the full allowed
range, $-1$ to $1$, and values for $B^{+0}/B^{+-}$, $B^{00}/B^{+-}$
and $\adir00$ in the specified range in the scenario. A total of
$10^5$ sets of values are generated. If a given set of values (i)
reproduces the measured value of $2\alpha$ in the allowed range (a) or
(b) [Eq.~(\ref{alphachoices})] using isospin, and (ii) gives $r^2$ in
the allowed theoretical range [Eq.~(\ref{rbound})] for $\tnp = 0$,
then it is consistent with the SM. If not, we conclude that new
physics is present.

\begin{figure}
\center
\rule{2cm}{0.2mm}\hfill \rule{2cm}{0.2mm}
\psfig{figure=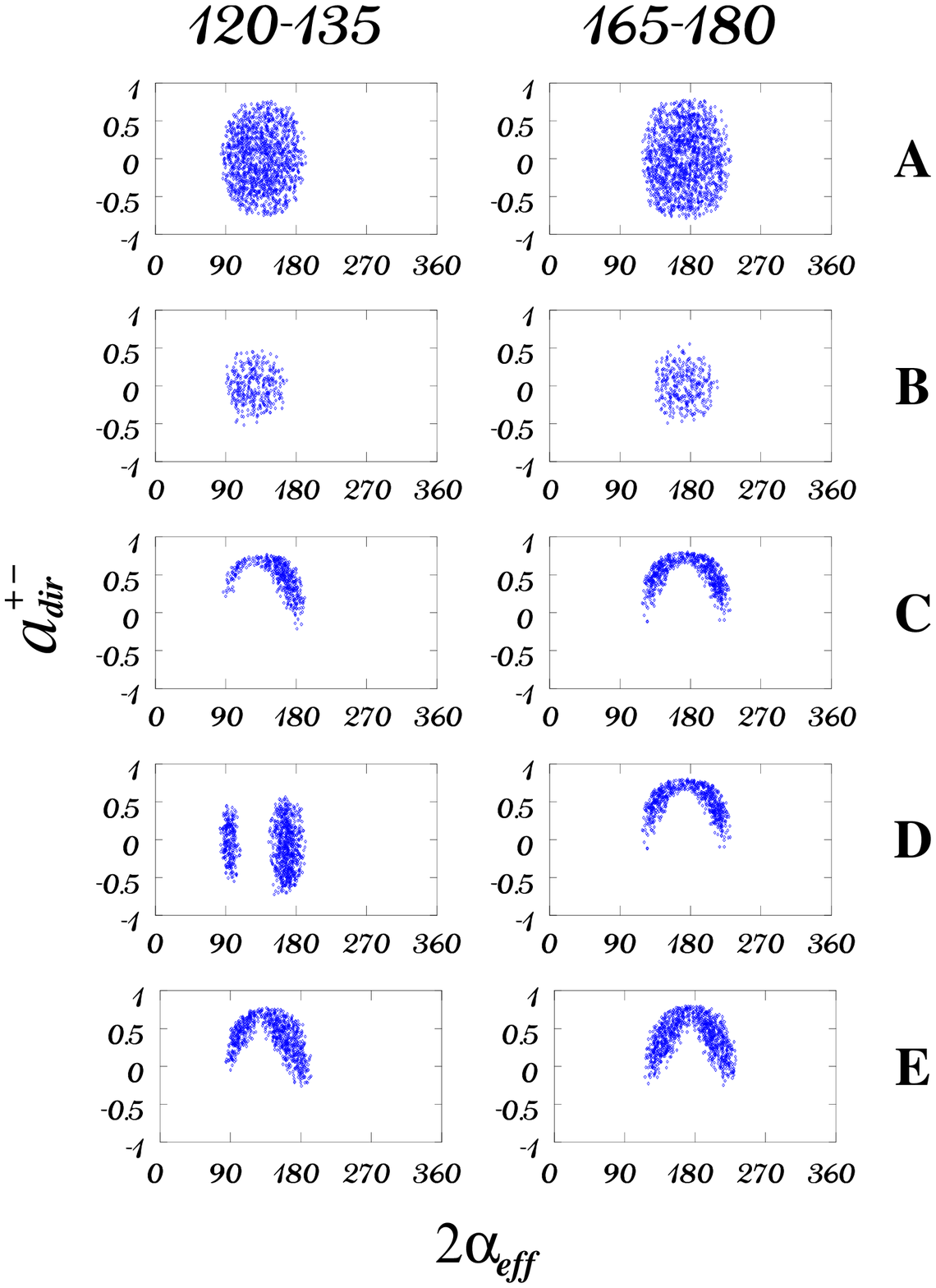,height=4.0in}
\rule{2cm}{0.2mm}\hfill \rule{2cm}{0.2mm}
\caption{The region in $2\alphaeff$--$\adirpm$ space which is
  consistent with the theoretical prediction for $|P/T|$
  [Eqs.~(\protect\ref{r}),(\protect\ref{rbound})]. In addition to the
  measurement of $\bd(t)\to\pi^+\pi^-$, it is assumed that information
  about $B^+\to\pi^+\pi^0$ and $\bd/\bdbar \to \pi^0\pi^0$ is
  available. For this latter information, the five scenarios of Table
  \ref{casetable} are considered from top (Case A) to bottom (Case E).
  In all cases, $2\alpha$ is allowed to take a range of values, given
  above each of the two columns of figures. In all figures, the
  $x$-axis is $2\alphaeff$ and the $y$-axis is $\adirpm$.}
\end{figure}


The results are shown in Fig.~1. The dark regions correspond to those
values of $\adirpm$ and $2\alphaeff$ which are consistent with the SM.
As one can see from these plots, there is a lot of white space. That
is, in each scenario, there is a large region of
$\adirpm$--$2\alphaeff$ parameter space which corresponds to new
physics. For example, consider scenario B. Only about $1/8$ of the
entire parameter space is consistent with the SM. Thus, even if our
knowledge of $B^{+0}/B^{+-}$, $B^{00}/B^{+-}$ and $\adir00$ does not
improve in the future, we have a good chance of seeing new physics,
should it be present, through the measurement of $\bd(t) \to \pi^+
\pi^-$ alone, along with an independent determination of $2\alpha$.

To be fair, things are not quite this easy. The presence of discrete
ambiguities in the extraction of $2\alphaeff$ and in the calculation
of $2\alpha$ via isospin can complicate matters. These complications
can be minimized if we have a variety of independent determinations of
$2\alpha$, as described earlier. For more details, we refer the reader
to Ref.~9.

To summarize: while the presence of physics beyond the SM in the $b\to
s$ FCNC is relatively easy to establish, the same is not true of the
$b\to d$ FCNC -- one always needs to add some theoretical information.
In this talk we have described how to use the $B\to\pi\pi$ system to
probe new physics in the $b\to d$ FCNC. Here one uses a prediction of
the ratio $P/T$ as the theoretical input. As we have seen, there is a
large region of parameter space where new physics can be found. Note
that the measurements of $B^+ \to \pi^+ \pi^0$ and $\bd/\bdbar \to
\pi^0\pi^0$ are not needed if $2\alpha$ can be obtained independently.
Ideally, we will have information about $2\alpha$ from both
independent sources and an isospin analysis.

\section*{\bf Acknowledgments}

We thank the organizers of BCP4 for an exciting, fascinating
conference. The work of D.L. was financially supported by NSERC of
Canada.

\end{document}